\documentclass[aip,pof]{revtex4-1}
\draft
\usepackage{bm}
\usepackage[T1]{fontenc}
\usepackage[latin1]{inputenc}
\usepackage{eso-pic}
\usepackage{color}
\usepackage{rotating}
\usepackage{amsmath}
\usepackage{amssymb}
\usepackage{natbib}
\usepackage{pstricks}
\usepackage{psfrag}
\usepackage[toc,page]{appendix} 
\usepackage{booktabs}
\usepackage{fullpage}
\usepackage{balance}
\usepackage{mathrsfs}
\usepackage{latexsym}
\usepackage{wasysym}
\usepackage{fix-cm}
\makeatletter
  \AddToShipoutPicture*{
    \setlength{\@tempdimb}{.35\paperwidth}
    \setlength{\@tempdimc}{.66\paperheight}
    \setlength{\unitlength}{1pt}
   \put(\strip@pt\@tempdimb,\strip@pt\@tempdimc){
      \makebox(200,-250){\rotatebox{60}{\textcolor[gray]{0.88}{\fontsize{5cm}{5cm}\selectfont{Preprint}}}}
    }
}
\makeatother

\begin{document}
\title{Nonlinear Dynamics of Coiling in Viscoelastic Jets}
\author{ T. S. Majmudar},
\affiliation{Hatsopoulos Microfluids Laboratory, Department of Mechanical Engineering, Massachusetts Institute of Technology, Cambridge, MA 02139, USA}
\author{ M. Varagnat},
\affiliation{Hatsopoulos Microfluids Laboratory, Department of Mechanical Engineering, Massachusetts Institute of Technology, Cambridge, MA 02139, USA}
\author{ W. Hartt},
\affiliation{Corporate Engineering Technologies Lab, The Procter and Gamble Co, West Chester, OH 45069}
\author{G. H. McKinley}
\affiliation{Hatsopoulos Microfluids Laboratory, Department of Mechanical Engineering, Massachusetts Institute of Technology, Cambridge, MA 02139, USA}

\date{\today}

\newcommand{\g}{\ensuremath{\gamma}}			
\newcommand{\gdot}{\ensuremath{\dot{\gamma}}}	
\newcommand{\str}{\ensuremath{\tau}}			
\newcommand{\sv}{\ensuremath{\eta}}			
\newcommand{\zsv}{\ensuremath{\eta_{0}}}		
\newcommand{\lam}{\ensuremath{\lambda}}		
\newcommand{\lambr}{\ensuremath{\lambda_{\rm br}}}		
\newcommand{\lamrep}{\ensuremath{\lambda_{\rm rep}}}		
\newcommand{\emod}{\ensuremath{G_{N}^{0}}}	
\newcommand{\emodp}{\ensuremath{G_{\rm P}}}	

\newcommand{\gcomp}{\ensuremath{G^{*}}}		
\newcommand{\gp}{\ensuremath{G^{\prime}}}		
\newcommand{\gpp}{\ensuremath{G^{\prime\prime}}}
\newcommand{\fns}{\ensuremath{N_{1}}}		
\newcommand{\sns}{\ensuremath{N_{2}}}		
\newcommand{\fnc}{\ensuremath{\Psi_{1}}}		
\newcommand{\snc}{\ensuremath{\Psi_{2}}}		

\newcommand{\rads}{\ensuremath{\rm rad\cdot s^{-1}}}	
\newcommand{\rs}{\ensuremath{\rm s^{-1}}}		
\newcommand{\pas}{\ensuremath{{\rm Pa}\,{\rm s}}}		

\newcommand{\da}{\ensuremath{a}}			
\newcommand{\z}{\ensuremath{h}}			
\newcommand{\e}{\ensuremath{\epsilon}}			
\newcommand{\ui}{\ensuremath{U_{\infty}}}			
\newcommand{\vi}{\ensuremath{V_{\rm ref}}}			

\newcommand{\cpy}{CPyCl/NaSal}				
\renewcommand\floatpagefraction{.9}

\begin{abstract}

Instabilities in free surface continuous jets of non-Newtonian fluids, although relevant for many industrial processes, remain less well understood in terms of fundamental fluid dynamics. Inviscid, and viscous Newtonian jets have been studied in great detail; buckling instability in viscous jets leads to regular periodic coiling of the jet that exhibits a non-trivial frequency dependence with the height of the fall. Very few experimental or theoretical studies exist for continuous viscoelastic jets beyond the onset of the first instability. Here, we present a systematic study of the effects of viscoelasticity on the dynamics of free surface continuous jets of surfactant solutions that form worm-like micelles. We observe complex nonlinear spatio-temporal dynamics of the jet and uncover  a transition from periodic to doubly-periodic or quasi-periodic to a multi-frequency, possibly chaotic dynamics.  Beyond this regime, the jet dynamics smoothly crosses over to exhibit the "leaping shampoo effect" or the Kaye effect. This enables us to view seemingly disparate jetting dynamics as one coherent picture of successive instabilities and transitions between them. We identify the relevant scaling variables as the dimensionless height, flow rate, and the elasto-gravity number and present a regime map of the dynamics of the jet in terms of these parameters.
\end{abstract}
\maketitle

\section{Introduction}
Free surface continuous jets of viscoelastic fluids are relevant in many industrial processes like fiber spinning, bottle-filling, oil drilling etc. In many of these processes, an understanding of the instabilities a jet undergoes due to changes in fluid parameters like Reynolds number or Deborah number is essential from process engineering point of view. From a fundamental fluid dynamics point of view, jets of viscoelastic fluids present many challenging problems, which remain unresolved; the addition of elastic effects on the dynamics of viscous jets is one such challenging problem. Viscous free surface flows, especially jets, have been extensively studied in many different contexts. From the earliest studies of breakup of viscous jets by Rayleigh \cite{Ray} and Taylor \cite{GIT}, free surface jets have provided spectacular phenomena like capillary breakup \cite{EggersRMP}, and the coiling instability \cite{CM-1}. These investigations have enhanced our understanding of the physics of free surface viscous flows by employing experiments, analytical treatments, and numerical modeling. 

Viscoelastic free surface flows add another dimension to the problem via elasticity of the fluids and provide equally stunning effects in free surface flows. For example, dripping of a polymeric fluid gives rise to beads-on-a-string structure due to elasto-capillary thinning \cite{oliveira}. While the beads-on-a-string phenomenon occurs at very low flow rates, at high flow rates and under electric fields a different set of problems occur as in the fiber spinning process \cite{GIT-2}.

In this paper, we investigate the coiling instability of buckled \emph{viscoelastic} jets and study the effects of rheology on the dynamics in fluids that form worm-like micelles. The problem of buckled viscoelastic jet has not been studied experimentally so far. This challenging problem is of interest not only for industrial applications, but also for understanding certain geophysical flows \cite{GT}.


The first study of the coiling instability in viscous jets was done by Barnes and Woodcock in 1958 \cite{Barnes-1}, who named the instability the ``rope-coil effect''.  The basic observations regarding the coiling instability were outlined in this paper. This study was followed by a second study, where the height  dependence of the coiling frequency was measured \cite{Barnes-2}. These two papers described a seemingly simple, liquid column buckling instability, which can be observed easily in everyday situations like pouring honey on toast. The problem of buckled liquid jets and sheets has since been investigated by a number of authors, with increasing experimental accuracy and sophisticated modeling. The next detailed investigation of this problem was carried out by Cruickshank and Munson in a series of papers \citep{CM-1,CM-2,CM-3,CM-4,CM-5}. These papers established a rigorous quantitative foundation for the problem of axi-symmetric, and planar viscous jet bucking.  The principal findings in these studies were the identification of critical height and limiting Reynolds number for the coiling instability. It was found that a viscous jet undergoes buckling only if the Reynolds number is less than 1.2, making it a low Reynolds number phenomenon. It was also found that the critical aspect ratio at which the buckling instability appears is $H/d = 7.7$ for axi-symmetric jets, and $H/d = 4.8$ for planar jets. Here $H$ is the drop height and $d$ is the nozzle radius \citep{CM-1, CM-5}. The additional observation relates to the variation of frequency of coiling with the aspect ratio. A simple model for this variation was put forth based on energy dissipation arguments \citep{CM-3,CM-4}. On the theoretical front, one of the first investigations that dealt with describing coiling and folding in viscous jets and sheets were by Tchavdarov et al. \cite{Yarin-1} and Yarin and Tchavdarov \cite{Yarin-2}. Using perturbation theory,  these authors were able to model the jet beyond the buckled state and obtain good agreement with the observations of Cruickshank and Munson. 

A resurgence of interest in this problem began as a result of the study by Ryu et al. \cite{maha-1}, who revisited the problem of coiling in viscous jets and combined experiments and theoretical arguments to put forth scaling laws for the dimensionless frequency and dimensionless amplitude as a function of fluid parameters and kinematic variables \cite{maha-2}. This study was extended by Maleki et al. \cite{Maleki-1} and Habibi et al. \cite{Habibi-1}, in which they obtained scaling laws for the frequency variation in viscosity dominated regime, viscous-gravitational regime, inertio-gravitational  regime, and inertial regime. These scaling laws were obtained by balancing dominant forces acting in each regime. In the transitional inertio-gravitational regime, Ribe et al. found the existence of multiple steady states, wherein the coiling occured at two different frequencies \cite{Ribe-1}. These frequencies were seen to be eigensolutions to the whirling slender string model proposed in this paper. Although the theoretical solutions admitted more than two frequencies, the experiments could capture only two of those frequencies, and the branch connecting these two frequencies was found to be unstable. Ribe also proposed a general theory of coiling and folding in viscous threads and sheets \citep{Ribe-2,Ribe-4}. 
Several other related problems of interest have been studied since then; for example, the coiling of elastic ropes \cite{Habibi-2}, instabilities in dragged viscous threads \cite{Ribe-3}, and the meandering instability of viscous threads \cite{Morris}. Lastly,  a new one dimensional model has been proposed by 
Nagahiro and Hayakawa \cite{NH}, where they obtain many of the observed features of viscous coiling from a simplified one dimensional flow model.

In contrast to this extensive body of work on the buckling instability of viscous jets and sheets, no such experiments have been carried out for viscoelastic jets. On the other hand, there have been numerous  theoretical treatments of slender viscoelastic jets, though none of these studies deal with buckled jets. 
An early work in this area is by Matovich and Pearson \cite{Matovich}, where the authors treated a viscoelastic jet in fiber-spinning geometry. Koesser and  Middleman \cite{MM} treated slender viscoelastic fibers to assess the stability of such fibers against breakup. A model of viscoelastic fibers as one dimensional slender body has been achieved in a series of papers \citep{BFW-1,BFW-2,BFW-3}, where the authors reduce the three dimensional problem to an effectively one dimensional problem, and obtain a set of coupled nonlinear ordinary differential equations for the system. 
 
 At the higher Reynolds number, viscoelastic fluids, especially surfactant fluids show a remarkable phenomenon discovered first by A. Kaye \cite{AK}. He observed that when a solution of polyisobutylene in decalin is poured from a large height on to a plate, the liquid stream forms a mound, now known to be a mound of coils, but intermittently jumps vertically or sideways to large distances. These ``leaps'' were considered to be a type of ``recoil'' for the liquid stream. The next study to address this issue was by Collyer and Fisher \cite{CF}, who suggested that shear-thinning was an essential property for fluids to exhibit this ``leaping effect''. In addition, they suggested that any polymeric fluid,  which was elastic, ``pituitous", and highly shear-thinning would show the Kaye effect. Since then, it is known that many shampoos or liquid detergents show the same effect. 
 
 This problem was examined again, after a gap of thirty years, in a recent study by Versluis et al. \cite{Vers}, who put this problem on a quantitative footing. They examined the effect using high speed imaging and found that indeed shear-thinning is the most essential property of fluids that exhibit the Kaye effect. The authors also proposed that elasticity played no role in the origin or stability of the effect. A simple model based on energetics was put forward, with a suitable function describing the decrease of viscosity with shear rate and good agreement was found between experimental observations of the velocity of the leaping jet and those predicted by the model. They also observed that a minimum flow rate and a critical height were needed to establish the leaping shampoo effect.  

 The dynamics of free surface viscoelastic jets cover a wide array of phenomena. Many questions arise as one crosses over from Newtonian to non-Newtonian fluids; how does fluid elasticity change the jetting dynamics? What influences do shear and extensional rheology of the fluid have on the behavior of the jet? Here we address some of these issues within the context of buckling of viscoelastic jets and their subsequent dynamics. 
We report results of a systematic study of continuous jets of shear-thinning worm-like micellar fluids of varying viscosities and elastic relaxation times. This enables us to compare the role of viscosity and elasticity on the jetting dynamics. We completely characterize the rheology of the fluids and the jetting dynamics is observed by digital video imaging. We identify relevant parameters and define the problem in terms of dimensionless variables. We cast different jetting regimes in terms of a regime map that combines kinematic, geometric, and rheological parameters. 

\begin{table}[hbtp!]
\caption{Overview of test fluids used in the present study and relevant physical parameters at 22$^{\circ}$C. Here $\eta$ is the stated viscosity in the limit of zero shear rate, $\lambda$ is the elastic relaxation time, and the Elasto-gravity number is given as: 
$E_{g} = \lambda {\left(\frac{\rho g^{2}}{\eta}\right)}^{1/3}$}
\begin{center}
\begin{tabular}{|c|c|c|c|c|}
\hline
Fluid&$\eta$&$\lambda$&Rheology&Elasto-gravity number~($E_g$)\\
\hline
S1&$~9.3~$&$~0.03~$&~Shear thinning~&$~0.7~$\\
\hline
S2&$~16.5~$&$~0.1~$&~Shear thinning~&$~1.9~$\\
\hline
S3&$~34.8~$&$~0.16~$&~Shear thinning~&$~2.4~$\\
\hline
\end{tabular}
\end{center}
\label{tab:fluid properties}
\end{table}

\section{Model Fluid}
In order to investigate the effects of viscosity and elasticity on the jet dynamics, we need a model fluid whose viscosity and elastic relaxation time 
can be varied in a systematic fashion. To this end, we choose as our model fluid, the surfactant solution of Sodium Lauryl Ether Sulphate (SLES). 
The choice of this fluid as a model fluid stems from the fact that SLES is the main surfactant in many shampoos and household cleaners. These fluids are shear thinning and they form worm-like micelles.  
By adding appropriate amounts of Sodium Chloride (NaCl), the viscosity and the elastic relaxation time of the resulting fluid can be varied systematically. 

Three different test fluids were prepared in this manner by adding 
$11.4~ {\rm gm} $, $13.4~ {\rm gm}$, and $17.8~ {\rm gm}$ of NaCl to $500~ {\rm ml}$ of SLES. The salt was added and the fluid was placed on a rotating mixer for a period of 24 hours, and then allowed to sit for another 24 hours before use. The resulting fluids were 
adjusted to be $250~ {\rm mM/L}$ (S1), $300~ {\rm mM/L}$ (S2), and $400~ {\rm mM/L}$ (S3) solutions of SLES and NaCl. The properties of these model fluids are listed in Table $1$. It can be observed from Table $1$ that adding salt to the surfactant solution increases both the viscosity and the elastic relaxation time 
($\lambda$). The increase in elastic relaxation time signifies that the fluid is more elastic. The viscosities and the elastic relaxation times increase linearly with salt concentration as shown later.

\section{Experimental Methods}

\begin{figure}[bthp!]
\centering
\vspace{10pt}\
\includegraphics[width=400pt]{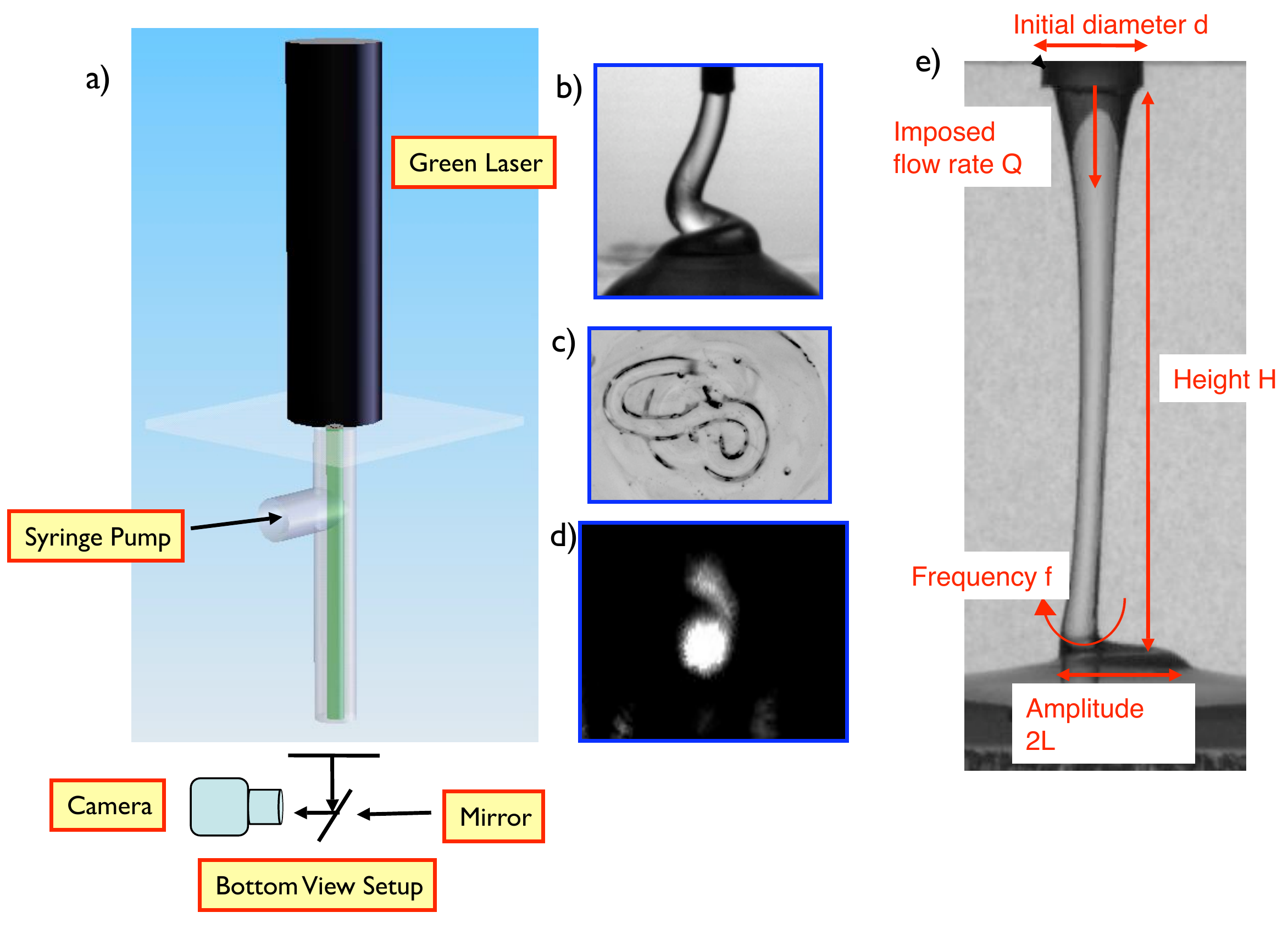}
\caption{Schematic diagram of the experimental setup and different viewing configurations. a) Schematic diagram of the bottom-view setup shows a laser light going through a T-junction. The other end of the the junction is the fluid inlet. The junction is mounted on a linear stage. The fluid falls on a plexiglas plate. b) An image of the jet in the side-view configuration, where the jet is backlit with halogen lamp and the camera is placed perpendicular to jet direction. In this configuration, the laser lighting is not employed. c) An image from the bottom-view setup with halogen lighting; the light falls onto the plexiglas plate, the mirror at $45^\circ$ projects the jet outline and the trajectory in two-dimensional space onto the camera. d) An image from the bottom-view configuration with laser light through the jet. The tip of the jet makes a spot, which moves along with the jet. e) The profile of the jet with the relevant parameters.}
\label{fig:fig1}
\end{figure}

The jetting dynamics was observed using digital video imaging under three different configurations. Figure~\ref{fig:fig1} describes the setup and the experimental arrangement with an example of image captured using each view configuration. The reason for imaging the jet dynamics using three different view-configurations is to be able to resolve the complex spatio-temporal dynamics. Unlike Newtonian viscous jets, where the principal complexity is in the frequency of coiling with increasing height, in non-Newtonian jets the dynamics is complex both spatially and temporally. 

In Fig.~\ref{fig:fig1}(a), we show the schematic diagram of the experimental setup. The fluid is pumped via a syringe pump (Stoelting). The fluid is pushed at a specified flow rate with an accuracy of $0.01~ {\rm ml/min}$. The fluid is stored in standard disposable plastic syringes (BD), with a capacity of either $10~ {\rm cc}$, or $60~ {\rm cc}$. The fluid then passes through Tygon tubing of fixed length ($\sim 18~$inches) for each experiment. The other end of the Tygon tubing has a steel nozzle of length   
1 inch and diameter, $d=1.2~ {\rm mm}$, attached to it. The fluid exiting the nozzle falls on a Plexiglas plate. The nozzle is fixed to a linear stage (Velmex) in vertical orientation. This allows us to vary the height of fall precisely with a resolution of $0.1~ {\rm mm}$. The height of fall in all experiments was varied between $0.5~ {\rm cm}$ and $20~ {\rm cm}$. 

The three different imaging configurations give three different views of the jet. The first case is the traditional side view. In this configuration a digital camera (BlueFox) is placed in a direction perpendicular to the direction of the flow of the jet, and the jet is backlit with a 500 W halogen lamp, with a diffuser screen placed between the light and the jet. The profile of the jet in this configuration is shown in Fig.~\ref{fig:fig1}(b). The advantage of this view is that the onset of coiling and the jet diameter are readily observable. The frequency of coiling at the onset can also be easily calculated using frame counting. 

As the jet dynamics becomes more complicated, for example when the trajectory of the jet follows a complicated two dimensional path, the side view provides insufficient information regarding the jet dynamics. In order to extract this missing information we employ two more view configurations; both involve imaging the jet from the bottom but in one case we use the diffused halogen illumination, and in the other case we use a beam of green laser passing through the jet. The idea of using a laser through a jet was first employed by Versluis et al. \cite{Vers}. 

The bottom view configuration with the laser light is shown in Fig.~\ref{fig:fig1}(a). The green laser light is placed vertically, and the light enters a T-junction. The end of the T-junction collinear with the laser is sealed off using a thin transparent acrylic sheet. The fluid from the syringe enters the second opening of the T-junction, perpendicular to the direction of the laser. The fluid then moves downwards, and exits out of the remaining end of the T-junction. The beam of laser passes through the fluid, and acts as a light-guide. In the bottom view, a mirror is placed at $45^{\circ}$ below the plexiglas plate on which the jet falls. The point where the jet strikes the plate shows up as a bright spot, which moves as the jet moves in the {\it x-y} plane. This imaging technique works well for small amplitude and low frequency coiling or linear oscillations (folding). 

When the jet executes a complex two dimensional motion rapidly, the laser spot is unable to follow the jet in real time. In order to accurately capture the jet dynamics, we employ the second bottom view configuration, now with diffused light. The principle of the bottom view with diffused light is the same, except that instead of a bright point, as with a laser, one observes the edges of the whole fluid thread deposited on the plate, and its subsequent motion. In Figs.~\ref{fig:fig1}(c) and (1d), we show the images of the jet in the bottom-view configuration with diffused light, and laser light respectively. 
The digital videos are captured at frame rates ranging from 50 fps to 300 fps, giving us sufficient resolution to track the jet tip in two-dimensional space over time. 

\section{Experimental variables}


Figure~\ref{fig:fig1}(e) shows a side-view image of the jet with the relevant parameters. These parameters can be classified into two groups; fluid parameters, and kinematic parameters. Fluid parameters are the viscosity ($\eta$), elastic relaxation time ($\lambda$), density ($\rho$), and surface tension ($\sigma$). Kinematic parameters are the fluid flow rate $Q$, the height of fall $H$, acceleration due to gravity $g$, and the geometric parameter nozzle diameter $d$, which determines the aspect ratio for the problem. 

Amongst the fluid parameters, the density, and the surface tension do no change appreciably  for the three fluids. The surface tension is around $0.025~ {\rm N/m}$, and the density is $1.2~ {\rm kg/m^3}$. The values for the viscosity and elastic relaxation times are given in Table 1. The kinematic parameters varied systematically in the experiments are the flow rate, the height of fall, and the nozzle diameter. The nozzle diameter does not significantly alter the behavior of the jet except setting a scale for the aspect ratio. Any change in the behavior of the jet is a result of an interplay between viscous force, gravitational force, inertia and elasticity. The same interplay can also be cast in terms of relevant time scales. 

The behavior of a free surface viscoelastic jet involves at least three distinct time scales; the elastic relaxation time ($\lambda$), the viscous time scale ($T_v$), and the flow time of the fluid ($T_{f}$).     
The last two are given by:
\begin{eqnarray}
T_{v} = \frac{\eta d}{\sigma}\\
T_{f} = \frac{H \pi d^2}{4 Q}
\label{eqnarray:tscales}
\end{eqnarray}

From the balance of viscous and gravitational forces one obtains a length scale and a time scale for the problem: 
  \begin{eqnarray}
l_{gv} = {\left(\frac{\eta^2}{\rho^2 g}\right)}^{1/3}\\
\tau_{gv} =  {\left(\frac{\eta}{\rho g^2}\right)}^{1/3}
\label{eqnarray:ltscales}
\end{eqnarray}

Many dimensionless variables naturally arise in the problem either due to balance of forces or due to 
the time scales involved. In the following we will use $l_{gv}$ and $t_{gv}$ to obtain non-dimensional parameters relevant to our problem. The first important dimensionless variable is the Reynolds number ($Re$), which is the ratio of inertial forces to viscous forces, given as:
\begin{equation}
Re = \frac{4 Q \rho}{\pi d \eta} 
\label{equation:Re}
\end{equation}

The other variables can be cast into dimensionless forms using scalings used in viscous jet studies. For example the height and the flow-rate, two main control parameters, can be made dimensionless as follows:
\begin{eqnarray}
\epsilon = \frac{H}{d}\\
H^{*} =  H{\left(\frac{g \rho^2 }{\eta^2} \right)}^{\frac{1}{3}}\\
Q^{*} = Q {\left( \frac{g \rho^5}{\eta^5}\right)}^{\frac{1}{3}}
\label{eqnarray:adim-1}
\end{eqnarray}

The effects of elasticity can be captured by introducing a suitable dimensionless variable like the Deborah number $De = \lambda \dot{\gamma}$, where $\dot\gamma$ is the shear rate. It is to be noted here that the shear rate is not constant throughout the jet, especially in buckled jets. 

\section{Fluid Rheology}

\begin{figure}[bthp!]
\centering
\vspace{10pt}\
\includegraphics[width=400pt]{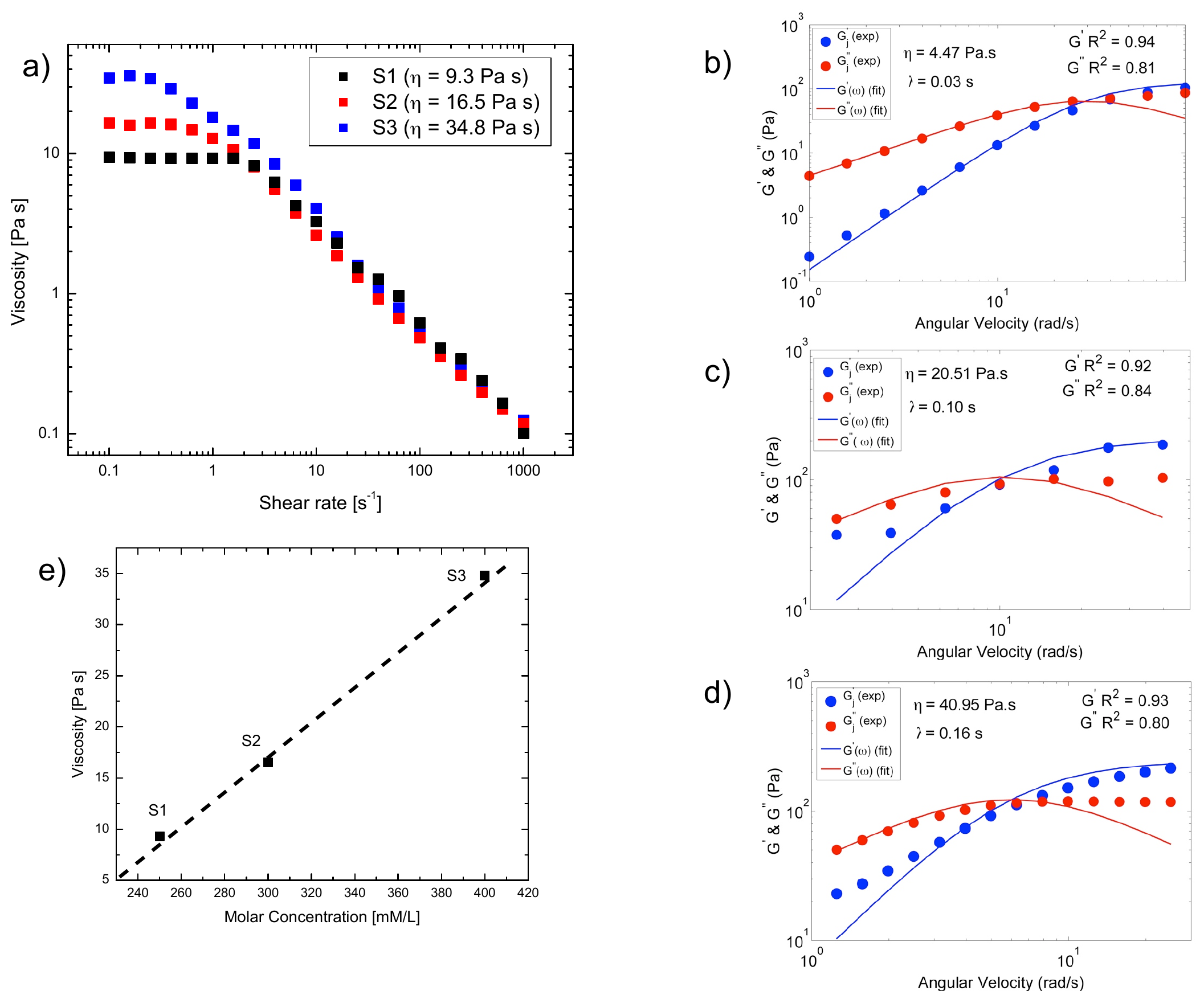}
\caption{Fluid Rheology: Viscous and elastic properties of the model fluids. a) Shear rheology of the three fluids showing variation of viscosity with shear rate. The initial plateau region gives zero-shear rate viscosity and each fluid exhibits strong shear-thinning, i.e. decrease in viscosity with increase in shear rate. The shear-thinning behavior begins at a shear rate of $\dot{\gamma} = 1~ s^{-1}$. b),c),and d) Small angle oscillatory shear response of the three fluids. Elastic (G'), and viscous (G") moduli are shown as a function of frequency for the fluid S1, S2, and S3 respectively. Each fluid exhibits a crossover point, beyond which, the elastic modulus becomes larger than the viscous modulus. The inverse of the frequency at the crossover point gives the elastic relaxation time of the fluid. A more rigorous approach entails fitting the moduli curves via a single mode Maxwell model, shown by solid lines. e) Viscosity variation as a function of the salt concentration;  viscosity varies linearly as a function of salt concentration.}
\label{fig:fig2}
\end{figure}

In order to decipher the effects of increasing viscosity and elasticity on the jet dynamics, it is essential to rigorously characterize the rheological properties of our model fluids. The data presented in Table 1 were obtained in the following manner.   

The rheological response of the model fluids was obtained using a conventional stress-controlled rotational rheometer ARG2 (TA Instruments) or strain-controlled rheometer ARES (TA instruments). Two types of rheological tests were conducted; steady shear-rate response, and small amplitude oscillatory shear (SAOS) tests. The steady shear-rate response gives the change in viscosity with shear-rate, and SAOS test gives the elastic and the viscous moduli for the fluids, from which an effective elastic relaxation time constant $\lambda$ can be extracted by fitting the data to a single relaxation time Maxwell model. 

First we describe the details of the steady shear-rate tests. The fluids were subjected to steady shear rate tests with shear rates varying from $0.1~ {\rm s}^{-1}$ to $1000~ {\rm s}^{-1}$, using a parallel plate geometry, with plate radii of $50$ mm, at a  gap height of 1 mm . The temperature was controlled using a Peltier plate at $22.5 ^{\circ}$C. In Fig.~\ref{fig:fig2}(a), we show the variation of viscosity with shear rate for the three fluids. Beyond a certain shear rate, the viscosity of each fluid begins to decrease, signifying a shear-thinning behavior. From this data, we obtain the zero shear-rate viscosities for each fluid, as stated in Table 1. 

The second set of tests performed were small angle oscillatory shear tests. These tests are aimed at obtaining the viscous ($G^{''}$) and elastic ($G^{'}$) moduli of the fluids. These tests were performed on the stress-controlled ARG2 rheometer, using cone-plate geometry. The diameter of the cone was 40 mm, and the cone angle was $2~^{\circ}$. Each fluid was placed within the cone-plate geometry, and oscillatory shear stress of $0.58~ {\rm Pa}$ was applied at frequencies ranging from 1 rad/s to 100 rad/s. We show the results in Figs.~\ref{fig:fig2}(b), (2c), and (2d) for the fluids S1, S2, and S3 respectively. As the frequency of oscillation increases, the elastic modulus $G^{'}$ increases. At a certain frequency, the elastic modulus crosses the viscous modulus. From this crossover point, a rough estimate of the elastic relaxation time can be obtained as the inverse of cross-over frequency. A more accurate method of obtaining relaxation time is to fit the single mode Maxwell model to experimental $G^{'}$ and $G^{''}$ curves. The elastic and viscous moduli are given by: 
\begin{eqnarray}
G^{'} = \frac{ \eta \lambda \omega^2}{1 + (\lambda \omega)^2}\\
G^{''} = \frac{\eta \omega}{1 + (\lambda \omega)^2}
\end{eqnarray}
where $\lambda$ is the relaxation time, $\omega$ is the frequency, and $\eta$ is the viscosity. 
The results of the relaxation times obtained via the fitting procedure are reported in Table 1. 
The effects of adding salt to SLES solutions is shown in Fig.~\ref{fig:fig2}(e). The viscosity increases linearly with salt concentration.

\section{Experimental Results}

To understand free surface flows of non-Newtonian fluids, it is important to study the behavior as the fluid parameters and kinematic parameters are changed. A useful representation would employ a three dimensional (3D) parameter space, where the axes are given in terms of a triplet of numbers, for example {Re, Wi, Ca}, or some other suitable set of non-dimensional parameters \cite{GHM-1}. 

In our case, for each fluid, we fix a flow rate and vary the height of fall to investigate the change in the dynamics of the flow. In this situation, we can employ a two dimensional (2D) representation of this 3D regime map. 
We determine the flow behavior in terms of dimensionless height, and dimensionless flow rate, and by comparing the jetting behavior for the three fluids with different viscosities and relaxation times, we can arrive at a representation in terms of other dimensionless  variables. 

It is instructive to recap the typical behavior of free surface viscous Newtonian jets. The initial jet dynamics is always stable, stagnation flow, where the jet spreads over the surface. The critical aspect ratio at which the jet undergoes buckling instability is a well-characterized \cite{CM-1}. The critical aspect ratio for a coiling transition is:
\begin{equation}
\epsilon = \frac{H}{d} =  4.8
\end{equation}
Subsequent to this buckling transition, the frequency of coiling changes as the aspect ratio increases. There is also an upper limit on the Reynolds number above which the jet returns to stagnation flow state. 
This critical Reynolds number is found to be 1.2. In all of our experiments, the Reynolds number as defined in Eq. 5 is much lower, within the range ${\rm Re} = [10^{-4} - 10^{-3}]$.
There exists another limiting behavior as the height of fall and the flow rate are varied - jetting to dripping transition \cite{zak}. This transition is determined by the flow time and the capillary breakup time, and is given by:
\begin{equation}
\epsilon = \frac{\eta Q}{\sigma d^2} 
\end{equation} 
The interesting dynamics we capture fall beyond these two limiting regimes.
For viscous Newtonian jets, the coiling frequency of the jet varies with height of the fall for a fixed flow rate. The coiling frequency can be obtained as a function of the height, the flow rate and the viscosity of the fluid. In different regimes, the scaling laws for the frequency are different, but for known flow rates and heights, coiling frequency for fluids with different viscosities can be ascertained using the scaling relations. For viscoelastic fluids, the situation is complicated by the introduction of elasticity, and its competition with viscous, gravitational, and inertial forces. Moreover, for shear-thinning fluids, as the height of fall is increased or the flow rate is varied, the fluid far away from the nozzle or in the coils may have a substantially different viscosity than the fluid exiting the nozzle. These considerations point to a need for quantifying the relative effects of elasticity, viscosity and gravity. One important parameter we use in this study is the elasto-gravity number, which compares these forces:

\begin{equation}
E_{g} = \lambda {\left(\frac{\rho g^{2}}{\eta}\right)}^{1/3}
\end{equation}
The elasto-gravity number increases as the effects of elasticity increase relative to viscous effects. In other words, $E_g$ increases whenever $\lambda$ increases or $\eta$ decreases for a finite value of $\lambda$. The elasto-gravity number has a clear lower limit; ${\rm E_g} = 0 $ for Newtonian viscous fluids. For the fluids used in this study, ${\rm E_g}$ is around $1.0$.     

\subsection{Instability in Viscoelastic Jets: General Observations.} 

\begin{figure}[bthp!]
\centering
\vspace{5pt}\
\includegraphics[width=480pt]{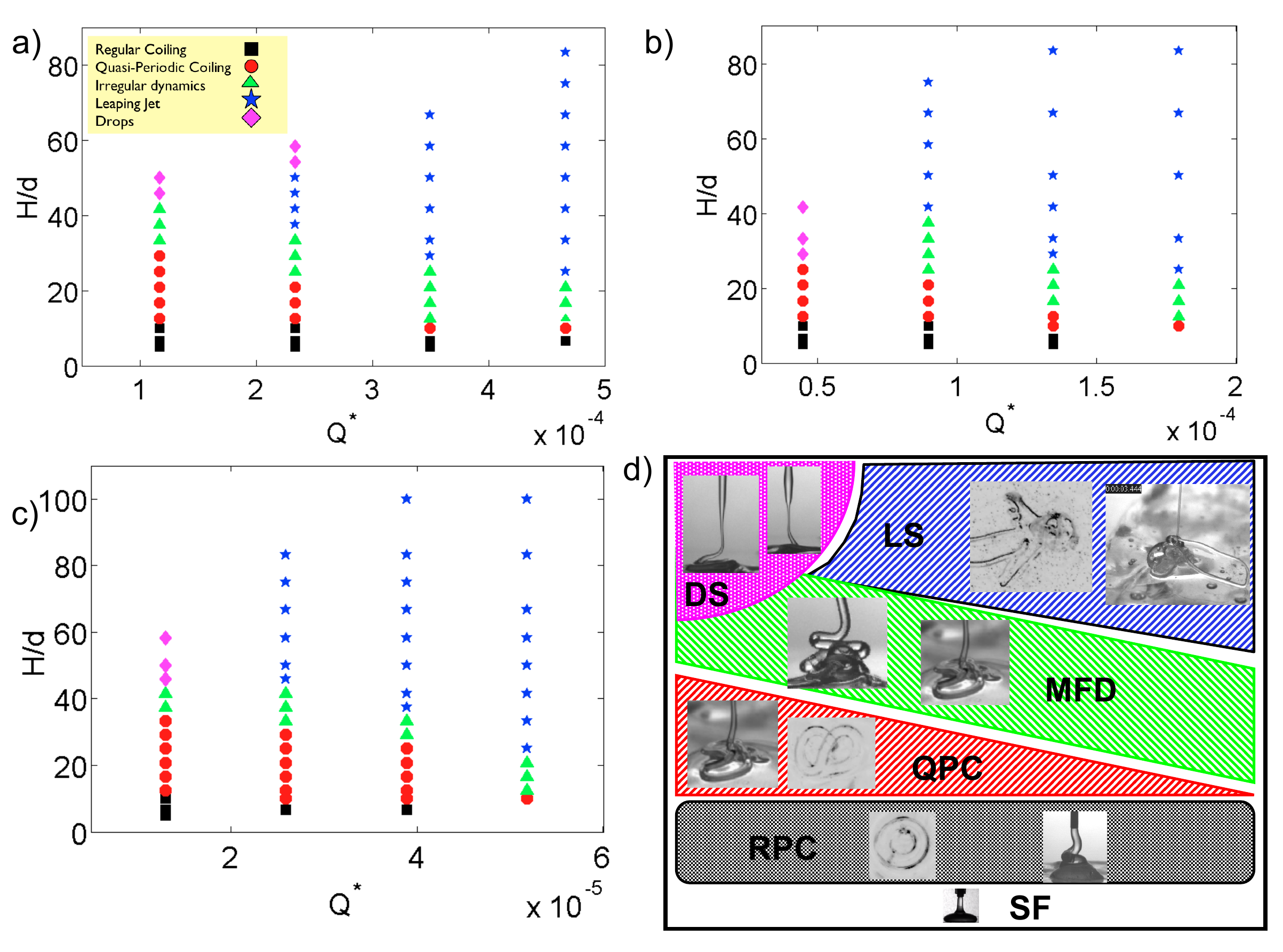}
\caption{Regime Map: Jet dynamics in ($H/D ~ - Q^*$) space. The behavior of the jet for the three fluids is shown at four different flow rates, as the height of the fall is varied. The legend is as follows: $\blacksquare$ represents the regular coiling state (RPC), the {\color{red} $\CIRCLE$} represents the doubly periodic or quasi-periodic coiling (QPC), {\color{green} \UParrow} represents multifrequency or irregular dynamics (MFC), {\color{magenta}$\blacklozenge$} represent dripping, and {\color{blue}$\bigstar$} star represent the ``leaping shampoo" state. a) Fluid S1; for the first two flow rates, the dynamics proceeds from RPC to QPC to MFC to dripping. For the latter two flow rates, the dripping state is replaced by an onset of ``leaping shampoo" state. b) Fluid S2; the dynamics is similar to fluid S1, except the dripping state is now accessible only for the lowest flow rate. The transition boundaries between the QPC state and MFC state are shifted compared to fluid S1. c) Fluid S3; the regime map shows the same generic features as for the other two fluids. d) Schematic regime map showing different flow behaviors. The pictures in the insets show the jetting dynamics either under side-view or bottom view. }
\label{fig:fig3}
\end{figure}

In this section, we describe some general observations of the jetting dynamics, which are generic and robust in that they are found for each fluid at different flow rates and heights. 
  For each experiment, a flow rate was chosen, the height of the fall was fixed. The flow was started and the jet dynamics was captured using a digital CCD camera. The height of the fall was then increased, and the same procedure repeated. 
  At each height, the flow was visualized using one or more view-configurations, and the process was repeated for a different flow rate. These entire set of measurements were performed for each of the three fluids. These observations were then combined, for each fluid, to produce a regime-map of the dynamical behavior at different flow rates and heights. 
  
Each fluid exhibits a similar pattern of dynamics as the height of fall is varied at different flow rates, though the dynamics itself is not always simple. This pattern of dynamics can be represented in a parameter space of scaled height, and scaled flow rate. These scalings can be chosen in multiple ways. 
As one set of scaled variables, we choose the scaled height $\epsilon = H/d$, and the scaled flow rate $Q^* = Q {\left( g \rho / \eta^5\right)}^{1/3}$. The dimensional flow rate is scaled with gravitational and viscous forces. The scaling analysis resulting from this choice of non-dimensional flow rate has been shown to be useful for viscous jets, though it leads eventually to a complicated set of relations for the frequency of coiling as a function of flow rate, height, and viscosity of the fluid. In order to compare our results to the viscous case, we retain this scaling of the scaled height and the scaled flow rate and report the regime map for each fluid with appropriate elasto-gravity number.  


In Figs.~\ref{fig:fig3}(a), (b), and (c), we show the regime maps of jet dynamics in $\epsilon - Q^{*}$ space for the three fluids S1, S2, and S3, respectively. The meanings of different symbols are explained in the figure legend. It is at once clear that the three fluids show the same type of behavior across flow rates and heights. Similar to the Newtonian case, the jet is stable below a certain critical aspect ratio, though it is not always possible to measure this state for long except for low flow rates, since the fluid heap clogs up the nozzle. For each figure, the four different $Q^{*}$ correspond to $Q = 1~ {\rm ml/min}$, $Q = 2~ {\rm ml/min}$, $Q=3~ {\rm ml/min}$, and $Q = 4~ {\rm ml/min} $, respectively. 

For each  flow rate a similar pattern of behavior is observed for each fluid as the height of fall is increased. This pattern can be simply described as follows:
\vspace{-12pt}
\begin{enumerate}
\item Stagnation Flow (SF)
\vspace{-12pt}
\item Regular Periodic Coiling (RPC)
\vspace{-12pt}
\item Doubly-Periodic or Quasi-Periodic Coiling (QPC)
\vspace{-12pt}
\item Multi-frequency Dynamics (MFD)
\vspace{-12pt}
\item Kaye Effect or the ``Leaping Shampoo Effect'' (LS)
\vspace{-12pt}
\item Dripping State (DS)
\end{enumerate}
The generic behavior for each fluid and the entire sequence of transitions is captured in Fig.~\ref{fig:fig3}(d). The pictures in the insets show either the side-view or the bottom-view for each kind of dynamic. 

We now briefly describe our definition of each of these states. Stagnation flow is the case when then jet is as yet unbuckled, the fluid hits the plate and spreads evenly on it. RPC state is one in which the jet having undergone buckling instability rotates around a fixed center, at a fixed frequency, either clockwise or anti-clockwise. This state is similar to the viscous coiling state in purely viscous jets.  What we call QPC here is a state in which the jet undergoes coiling motion with \emph{two} or \emph{three} principal frequencies. We note here that a subset of this type of motion has been termed {\emph{multiple coexisting states}} by Ribe et al. in their study of viscous jets \cite{Ribe-1}. The term QPC more accurately describes this behavior in the context of nonlinear dynamics exhibited by viscoelastic jets. The transition to a more complex dynamics is termed here as MFD, and it designates a coiling, folding, and meandering jet with multi-frequency behavior; the behavior is complex in both space and time making it a different class of motion than a whirling string exhibiting discrete multiple eigenfrequencies \cite{Ribe-1}. The next type of dynamics is the well known Kaye effect or more recently termed as the leaping shampoo effect. In this case, the jet exhibits inertial coiling at a high frequency with intermittent bursts of large amplitude displacements either in-plane or in a fully three dimensional space. 
These bursts may be of a single rope of jet or a fluid loop shooting out and away from the jet axis. 

Beyond the initial buckling instability, the jet first shows RPC. Upon increasing the height further, the jet undergoes a second transition to doubly-periodic or quasi-periodic coiling with two frequencies. Within this QPC regime, as the height of fall is increased, one more frequency is added to the dynamics and we observe behavior with three distinct frequencies. As the height is increased further, we obtain a spatio-temporally complex, multi-frequency dynamics. This dynamics could be chaotic within a small window of dimensionless height. On further increase of height, this chaotic looking dynamics, crosses over to  a different kind of dynamics, where the irregular meandering of the jet is replaced by the Kaye effect described earlier. 

For the lowest flow rate, for each fluid, we also obtain jetting-dripping transition above a certain height. 
In this state, while the jet does not breakup, it does not acquire steady flow conditions. Large beads of fluid tend to fall down along a very thin stretched fluid fiber, reminiscent of the well-known beads-on-a-string scenario \cite{oliveira}. 
In order to decipher the nature of 
these regimes and transitions between them, the motion of the jet has to be accurately determined. With a view to study the regime map in a more quantitative fashion, we make use of the bottom view imaging configuration and resolve the motion of the jet in XY plane over time. The data collected in the form of movies allow us to create the jet trajectory in space and over time.  

In the QPC regime and beyond, a simple frame counting approach proves to be insufficient. The measurement of frequencies of coiling is accomplished by tracking the jet tip, constructing the time series for the $X$ and $Y$ components of the motion, and computing the power spectra for each type of motion. The dominant frequencies in the power spectra correspond to different modes of motion like coiling or precession, provided the frequencies are independent and not simple harmonics of a fundamental frequency. 

In Figure 4, we show a series of jet trajectories and the power spectra of the $X$ and $Y$ components, for the RPC regime and beyond. The figures shown are for the fluid S2 at a flow rate of $Q = 2~ {\rm ml/min}$. The choice of analyzing the data in $X$ and $Y$ components, as opposed to more traditional $R$ and $\theta$ components stems from the nature of dynamics exhibited. Beyond the RPC regime, the jet may execute coiling as well as folding or linear oscillations. Moreover, the center of the rotation of the jet is also not fixed, but may execute periodic or aperiodic motion. This produces artificial discontinuities and additional errors in computing radial and angular coordinates as a function of time. 
In the following discussion, we look at each regime with the help of jet trajectory data and quantify the jet behavior in more detail. We will combine information from Figures 3 and 4 for each regime to get a better insight into the dynamics in each regime.

\begin{figure}[bthp!]
\centering
\vspace{5pt}\
\includegraphics[width=450pt]{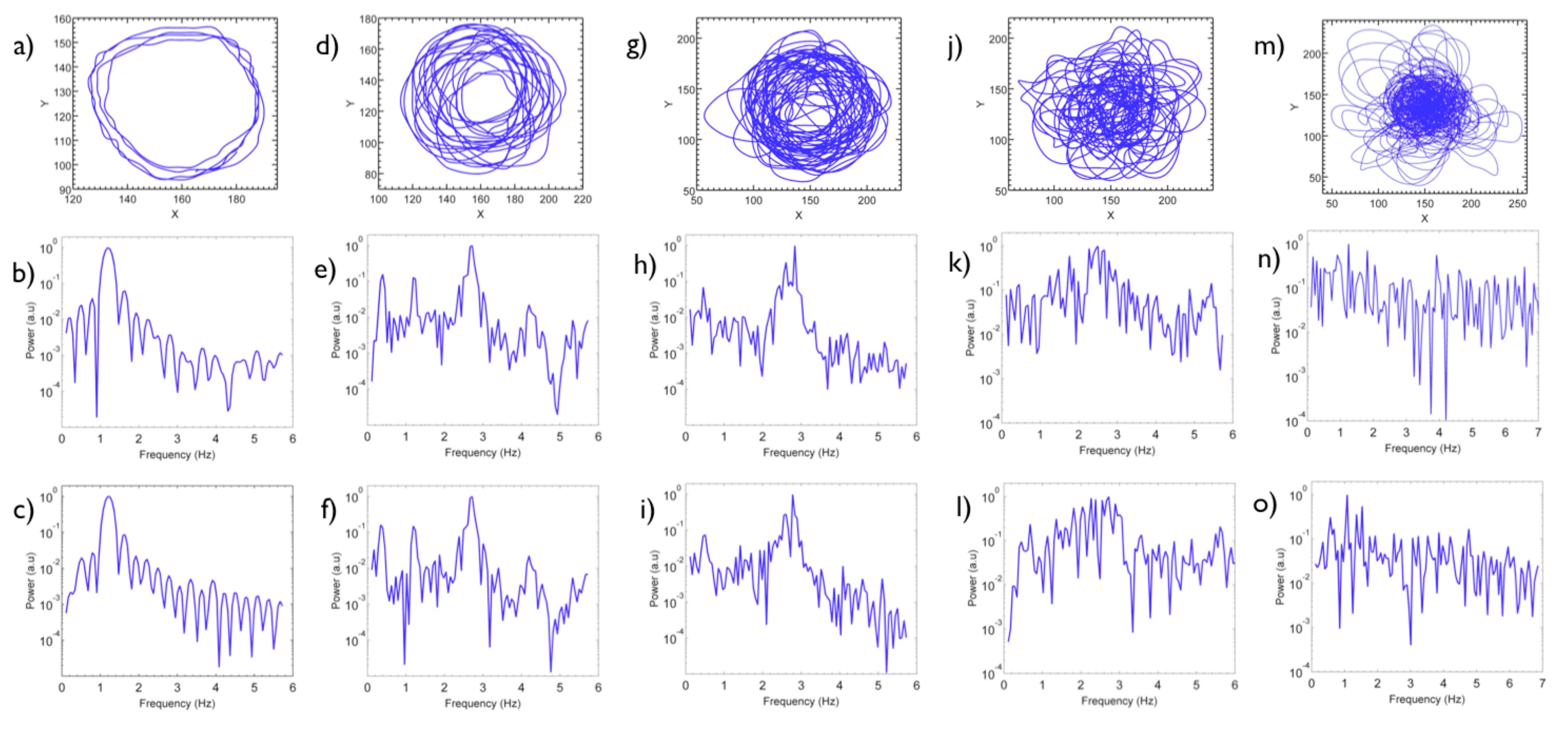}
\caption{Trajectory tracking and power spectra (semi-log scale): Top row shows the trajectories of the jets of fluid S2 at a flow rate of $Q ^* = 0.8 \times 10^{-4}$, for different heights. The power spectra of the $X(t)$ component are in the middle row, and the power spectra of the $Y(t)$ components are in the bottom row. a) Trajectory of the RPC state, at $H/d = 8.33$. b,c)  A single frequency $f_0=1.2 {\rm Hz}$ is clearly visible. d) The trajectory of the QPC regime at $H/d = 12.5$. The orbits are slightly elliptical and the jet exhibits coiling and precession. e,f) Power spectra of the $X(t),Y(t)$ components, respectively. The peak frequency at the previous height $f_0$ is still present but at a much lower amplitude. There are two additional frequencies, $f_1 = 2.72 {\rm Hz}$ and $f_2 = 0.40 {\rm Hz}$. g) The trajectory of another type of QPC dynamics showing coiling and folding states, which always appears after the rotation-precession dynamics. h,i) The power spectra of the components. 
j) The trajectory corresponding to the multi-frequency state. The jet exhibits complex coiling, folding and meandering dynamics giving rise to multiple frequency components. k,l) The power spectra corresponding to the $X(t)$ and $Y(t)$ components, respectively. The multi-frequency nature of the spectra is evident. m) The trajectory  at the onset of the leaping-shampoo state. Localized coiling and folding states are interrupted by large amplitude leaps in the $X-Y$ plane. n,o) The power spectra now have many frequencies and the power is distributed more uniformly over these frequencies.}
\label{fig:fig4}
\end{figure}

\vspace{-15pt}
\subsection{Regular Periodic Coiling}
 Beyond the stagnation flow, when the jet undergoes coiling transition, the observed coiling is highly regular, and periodic in space and time. Each ``fluid-rope'' falls exactly on top of the previous coil at fixed intervals, hence with a fixed frequency and amplitude (see Fig.~\ref{fig:fig3}(d), inset). Due to lower viscosities of the fluids used in these experiments as compared to earlier reported experiments in the literature, we do not observe coils stacking on top of one another in this regime; the time scale of fluid spreading on the plate is comparable to the time scale at which new fluid element arrives at the plate.  The RPC behavior is represented in Figs.~\ref{fig:fig3} (a), (b), (c) as black squares.
 
 The most significant feature of this state, in contrast to Newtonian jets, is the small window of scaled height in which RPC is observed. For viscous Newtonian jets, one observes a viscous regime in which the frequency of coiling decreases as the height is increased. Transition to gravitational regime reverses this trend and the frequency increases with fall height. For viscoelastic jets there appears to be a very small range of heights for which we observe RPC with a single frequency. This range is predominantly dependent on the dimensionless flow rate; higher the flow rate, lower this range. The maximum aspect ratio up to which RPC is obtained is for the fluid with lowest viscosity and elasticity ($\epsilon = 12.5$).  
The variation of RPC region on the elasto-gravity number also shows an inverse relation; the size of the RPC region decreases as the elasto-gravity number increases.  
 

The $X-Y$ trajectory of the RPC and the corresponding power spectra of each component are shown in Figs.~\ref{fig:fig4} a, b, and c respectively. The single frequency periodicity is revealed in both power spectra, with the peak frequency of $f_0 = 1.2~ {\rm Hz}$.    

 


\subsection{ Doubly-Periodic or Quasi-Periodic Coiling}  
The next stage in the dynamics occurs when the jet exhibits coiling with two or three frequencies. We call this regime the quasi-periodic regime for simplicity. The QPC regime is shown in Fig.~\ref{fig:fig3} as red circles. 
The range of heights for which the motion is quasi-periodic, again depends strongly on the flow rate; as the flow rate increases, for each fluid, the range of heights for which the motion is quasi-periodic, decreases. 

This type of doubly-periodic or quasi-periodic coiling comes about in a few different ways. 
Immediately following the transition from RPC, the jet undergoes both a rotatory motion and precession of the orbit around the center of rotation. This results in a characteristic {\it{rose curve}}, or a petal-like trajectory, as shown in  Fig.~\ref{fig:fig4}d (also see Supplementary Figure  2). The orbits are elliptical rather than circular.   In Figs.~\ref{fig:fig4}e and \ref{fig:fig4}f we show the power spectra of the $X(t)$  and $Y(t)$  components of the trajectory, respectively. The frequency of RPC, $f_0 =1.2~ {\rm Hz}$ is still present, but it is not the peak frequency. Instead, the peak frequency is $f_1 = 2.72~ {\rm Hz}$, and another frequency $f_2 = 0.40~ {\rm Hz} = f_0/3$ appears in the spectrum for $X(t)$. The ratio of the two frequencies is $f_1/f_2 = 6.86$. The faster frequency $f_1$ corresponds to coiling motion, and the slower one to precession of the jet. 
A second noteworthy feature  is the slight difference in the power spectra of the $X$, and the $Y$ components. The lower frequency for the $Y$ component is $f_2 = 0.34~{\rm Hz}$. The other two peaks correspond to $f_0 = 1.2~{\rm Hz}$, and $f_1 = 2.72~{\rm Hz}$. Two interesting relationships emerge in the spectrum for the $Y(t)$ component. The first relation is that $f_2 = f_1/8$, and the other is that the fundamental frequency present in RPC ($f_0$) is related to the two new frequencies by the relation $f_0 = (f_1 - f_2)/2$. 

This transition to quasi-periodicity is a novel feature of viscoelastic jets, and has not been observed in purely viscous jets. In addition, for each fluid, at every flow rate, the quasi-periodic behavior always begins with coiling and precession. It demonstrates that elasticity influences the emergence of {\emph{rotation-precession}} mode as the preferred mode once the RPC motion is unstable. The rotation-precession dynamics can be mapped to a trajectory on a torus. If the trajectory returns to its original starting position, after an integer number of turns around the torus, then the dynamics is phase-locked with two frequencies. On the other hand, a truly quasi-periodic trajectory does not return to its original location, since the ratio of frequencies is incommensurate. In this case, the trajectory densely fills the phase space. Experimental noise in the system makes it difficult to establish if the trajectories in Figs.~\ref{fig:fig4}d and \ref{fig:fig4}g are just phase locked or truly quasi-periodic, though the ratio of frequencies is not a simple integer ratio. 

Within the quasi-periodic regime, the petal-pattern dynamics transitions to a dynamics where both coiling and folding or linear motion are present. In this mode, the jet tip executes coiling motion interrupted by regular events of linear motion, sometimes forming a ``figure-of-eight''. Each time the jet undergoes a folding motion, the frequency of subsequent coiling switches between two or more steady state values, resulting in multiple frequencies associated with coiling and folding states. Figure 4g shows the trajectory of the jet, where circular coiling regions are present along with a few folding modes. The power spectra of the $X(t)$ and $Y(t)$ components of this state are shown in Figs.~ \ref{fig:fig4}(h) and (i), respectively. The spectra show three frequencies, two of which are coiling, and one for the folding state.  

This feature has been observed previously in viscous Newtonian jets, and it gives rise to multiple coexisting states in gravitational-inertial regime in the experiments of Maleki et al. \cite{Maleki-1}. These trajectories were also observed in experiments by \cite{Morris}, where viscous jets were allowed to fall on a moving belt. In their experiments the additional force is the drag on the fluid rope due to the moving belt. The authors in this study also point out that a weakly nonlinear theory may be required to explain these complex trajectory patterns. In the theoretical model for multiple steady states for viscous jets \cite{Ribe-1}, the two stable branches of solutions were connected by an unstable branch; correspondingly, only two frequencies were observed experimentally for a given height. In contrast, for viscoelastic jets, we have been able to observe three distinct frequencies in this regime. This suggests that introduction of elasticity stabilizes the previously unstable branches via an interplay of elastic, viscous, and gravitational forces. 


In addition to this mode, we have also observed a mode in which the jet precesses with a larger frequency and a smaller radius, and a lower frequency and larger radius, repeatedly. For each fluid, we observe two distinct frequencies at the same height. Again, this feature is also present in the viscous jets in the inertio-gravitational regime. For viscoelastic jets this mode is not the preferred multi-frequency mode in the inertio-gravitational regime. 




\subsection{Irregular Dynamics}  

As the height of fall is increased further, we see clear deviations from the Newtonian jet dynamics. 
In the Newtonian case, once inertial forces dominate the flow, the multiple coexisting states transition into regular coiling states at  very high frequencies. Instead,  for shear-thinning viscoelastic fluids we observe an increasingly irregular and possibly chaotic dynamics as the height of fall is increased further. 

The tip trajectory for this regime is shown in Fig.~\ref{fig:fig4}(j). It can be observed that for irregular dynamics, the trajectory though confined within a certain area, starts to occupy the area more uniformly. The dynamics now comprises of brief coiling states, interspersed with linear oscillations, and non-uniform meandering states. It should be emphasized that this is not a transient dynamic; the combination of coiling, folding, and meandering continues indefinitely, or in practice for as long as the jetting is observed.  
Figures~\ref{fig:fig4}(k), and \ref{fig:fig4}(l) show the power spectra of the $X(t)$ and $Y(t)$ components respectively for irregular dynamics. It can be clearly seen that the power is distributed more equally over a wider range of frequencies, all of which are not simple harmonics of some fundamental frequency.
Instead, we find that the peak frequencies have corresponding sub-harmonics. For example, in the spectrum for $X(t)$, we find that one of the peak frequency $f_0 = 2.5~ {\rm Hz}$ has a sub-harmonic $f_0 / 2 = 1.25~ {\rm Hz}$. The second frequency $f_1 = 2.27~ {\rm Hz}$ has a sub-harmonic of $f_1 / 5 = 0.45~ {\rm Hz}$. The third high frequency $f_2 = 2.665~ {\rm Hz}$ has no sub-harmonics or harmonics.  
The non-harmoniacally related multiple frequencies, and the subharmonics point to a subsequent parameter window of possibly chaotic behavior. 

Unambiguous determination of chaos in this system requires much longer time series, so that Lyapunov exponents can be calculated and strange attractors can be found. In addition, other dimensional measures, correlation functions, and return maps need to be found in order to characterize route to chaos in the system. Such a detailed characterization though currently under way, is beyond the scope of the present paper, and will be covered in a future publication.
The spatio-temporally irregular behavior is another novel feature of viscoelastic jets, not seen in purely viscous Newtonian jets. 

\subsection{Leaping Shampoo}
For viscous Newtonian jets, the inertia dominated regime is characterized by high frequency coiling regime, where the coils stack up on each other until the heap breaks and a new stack of coils takes its place. In the case of surfactant fluids, there is an additional, visually striking effect, known as the leaping shampoo effect, or the Kaye effect, first described by Kaye \cite{AK}. In this behavior, a jet of viscoelastic surfactant solution undergoes rapid coiling like a viscous jet, but intermittently exhibits large jumps away from the point of impact. A complete understanding of this effect including all the factors affecting this process has not yet emerged. A recent study \cite{Vers} gave a simplified model based on energy arguments, which describes the incoming jet causing a dimple in the surface of the pool of liquid, and beyond a critical height, results in an upward shooting jet. In this work, only shear thinning was seen as the factor responsible for the Kaye effect. Elasticity did not play a role in their model.  

There are two key rheological features of surfactant solutions; elasticity, and shear-thinning. Our investigations show that both have a role to play in the development and subsequent dynamics of the Kaye effect. We have verified (data not shown) that a fluid with constant viscosity ($\eta = 75~{\rm Pa~ s}$) and an elastic relaxation time of ($\lambda = 5.5~ {\rm s}$) like PS/PS Boger fluid {\emph {does not}} exhibit the Kaye effect.  This shows quite clearly that shear-thinning is essential for emergence of the Kaye effect,  consistent with the analysis by Vesluis et al. On the other hand, some shear-thinning but highly elastic surfactant fluids like aqueous solutions of Cetyl Pyridinum Chloride (CPyCl) {\emph {do not}} exhibit the Kaye effect \citep{MV-1,MV-2}. The CPyCl solutions we have tested typically have  a zero-shear viscosity $\eta = 15~{\rm Pa~s}$, and an elastic relaxation time $\lambda = 0.5~{\rm s}$. The elasto-gravity number for CPyCl solutions $E_g \approx 10$, whereas for the shampoo solutions in the present study, the elasto-gravity number $E_g \approx 1$. Thus, only surfactant fluids within a small parameter window of elasticity and shear thinning rheology exhibit leaping. This new finding suggests that elasticity does play a role in the development of the Kaye effect. 
More quantitatively, we can infer that the Kaye effect is most readily observed for shear-thinning fluids with elasto-gravity number $E_g \sim O(1)$. Furthermore, from the regime maps for each fluid  (Fig.~\ref{fig:fig3}), it is clear that the onset of the Kaye effect is also dependent on the flow rate, and is different for each fluid with different $E_g$. This implies that the critical height at which the Kaye effect begins is also a function of the viscoelasticity of the fluid.   

  Consistent with the observations by Versluis et al., we have observed two distinct modes of leaping; one in which the jet slips off the fluid surface and gets splashed more or less in-plane, but to a large distance compared to the size of the coiling region, and the other in which the jet penetrates the fluid bath, and shoots vertically up, which is the traditional Kaye effect. 
As the height of the fall is increased, the in-plane leaps of the jet due to slipping always precede the vertical leaps. The in-plane leaps of the incoming jet show no preferred angular orientation, and are in random directions with respect to the center of the jet. The direction of the leap being determined predominantly by the random slip direction of the incoming jet.

The trajectory of the jet for the leaping shampoo (LS) effect is shown in Fig.~\ref{fig:fig4}(m), where the leaps are predominantly in the XY plane. The dynamics in the center is still irregular at this stage with multiple frequency components. The jet shoots off from the center either as a loop of fluid, or as a single strand of fluid, which impacts the fluid bath a large distance from the center. These loops of fluid are clearly visible in the XY trajectory of the jet. 
The power spectra of the LS regime for the $X(t)$ and $Y(t)$ components are shown in Figs.~\ref{fig:fig4}(n), and \ref{fig:fig4}(o), respectively. The spectra consists of many frequency components. The highest frequencies  corresponds to the most rapid coiling motion around the point of impact. The other frequencies correspond to different coiling, folding, and leaping modes. 
  
On further increase of height, the jet exhibits vertical leaping, with very high frequency coiling around the center. From this point onwards, due to the resolution of the imaging, it was not possible to track the trajectory of the jet close to the center of the jet, although it is still possible to track the jet-leaps. On the other hand, the dynamics now is fully three dimensional and the bottom view only gives a projection of the jet motion, and not its accurate position. The leaping shampoo state continues to occur until the aspect ratio becomes so large that the jet diameter becomes less than the capillary length of the fluid. At this point, the jet is no longer steady, and we revert back to the dripping mode.  

\begin{figure}[bthp!]
\centering
\vspace{5pt}\
\includegraphics[width=480pt]{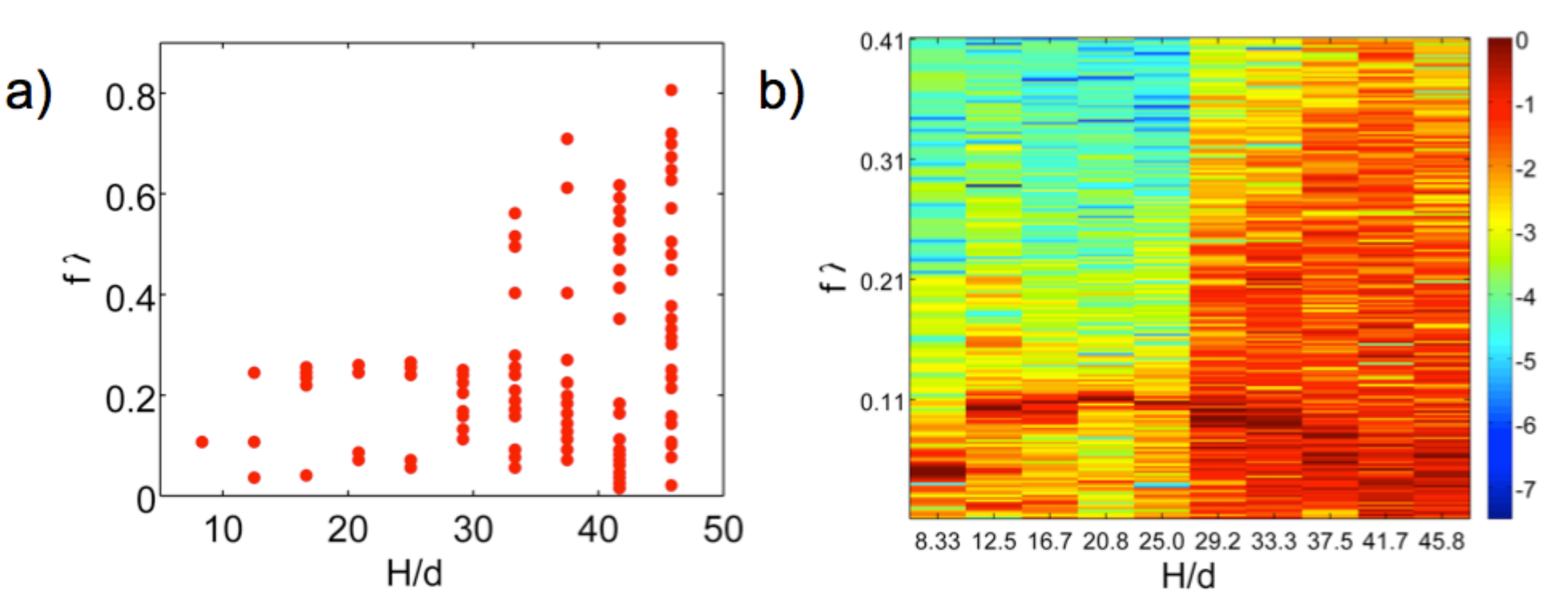}
\caption{Frequency space bifurcation diagram of fluid S2, at a flow rate of $Q^* = 0.9 \times 10^{-4}$. a) A plot of frequency scaled with the elastic relaxation time against the height scaled by the diameter of the nozzle. Each data point represents an independent frequency in the power spectrum of the $X(t)$ component of the trajectory. The first point is the single frequency RPC state, which bifurcates to a QPC state at the next height. At each successive height the peak frequencies change, and more frequencies get added until at the highest scaled height the spectrum has many frequency components. b) A different representation of the frequency bifurcation diagram in terms of an intensity map. Here the spectrum of $X(t)$ component is shown, with the intensities corresponding to the log of the power at each frequency. This representation reveals the relative power distribution amongst the frequencies. Again, the bifurcation of the frequencies at lower scaled heights, and the even distribution of the power at higher scaled heights is readily visible. }
\label{fig:fig5}
\end{figure}

\subsection{Frequency Transitions}

We focus here on one of the flow rates ($Q = 2~ {\rm ml/min}, Q^{*} = 8~\times~10^{-5}$), for the fluid S2.
From the power spectra of the $X(t)$ and $Y(t)$ displacements, the frequency components involved in the dynamics can be found. The zero frequency component is not considered in this process as it only gives the average displacement. For each height, the dominant frequency of coiling has the maximum amplitude. Typically, the dominant frequency corresponds to a coiling mode or a folding mode. Each independent frequency in the dynamics of the jet will also have its corresponding harmonics. This gives an additional measure for ascertaining the relevant frequencies. 

Beyond the quasi-static regime, when multiple frequencies are present, we find frequencies whose ratios are not simple integers or linear combinations of each other. We take these frequencies as independent frequency components of the spectrum. In addition, we find sub-harmonics of major frequencies in the spectrum. The sub-harmonics of a given frequency in the spectrum signifies period-doubling, period-tripling, or in general period-N oscillations (coiling or folding). The entire frequency information can be shown as a bifurcation diagram, which proceeds from zero frequency stagnation flow to period-1 dynamics (RPC), and transitions to quasi-periodic and possibly chaotic dynamics. In Fig.~\ref{fig:fig5}(a) we show the bifurcation diagram for the jet dynamics in terms of frequency scaled by the elastic relaxation time $\lambda$, and the height scaled by the nozzle diameter. The frequencies plotted are the independent frequencies and the subharmonics found for those frequencies. The frequency scaling used by Maleki et al. \cite{Maleki-1} in their studies of viscous jets is given by: 
\begin{equation}
f^* =  f{\left(\frac{\nu}{g^2}\right)}^{\frac{1}{3}}
\end{equation}
where, $f$ is the observed frequency, and $\nu$ is the kinematic viscosity. This frequency relates very simply with the frequency scaling used here via the Deborah number $De$ and the Elasto-gravity number $E_g$:
\begin{eqnarray}
De = f \times \lambda\\
f^* = \frac{De}{E_g}
\end{eqnarray}

The region before the first point in the plot corresponds to zero frequency stagnation flow. The onset of coiling is the first point, where the jet rotates with a single frequency; the regime we have called Regular Period Coiling (RPC). As the height increases further, the jet  undergoes quasi-periodic transition, and rotates with two distinct frequencies; one for coiling, and the other for precession. The quasi-periodic regime continues with addition of a frequency component as the height of drop is increased. Eventually, the jet executes coiling, folding, and meandering motion with multiple frequencies. 

In Fig.~\ref{fig:fig5}(b) we show the whole spectrum for the $X(t)$ component for each height as an intensity map. Here the intensity of the band scales with the logarithm of magnitude of the power at that frequency, normalized by the maximum power in the spectrum. The bifurcation in the frequency spectrum is clearly visible at the second height, which corresponds to $H/d = 12.5$, the onset of quasi-periodic transition. The intensity colormap clearly reveals the power distribution at various frequencies as the height of fall is increased. The power initially distributed within a narrow band of frequencies, gets more evenly distributed as the dynamics becomes more irregular at larger drop heights. Beyond the RPC regime, each higher peak frequency bifurcates into two closely spaced frequencies at the next height, and other frequencies corresponding to additional folding or meandering modes get added to the spectrum. This is the most detailed information that can be obtained from the present set of experiments, based on the resolution of the experiments. It shows that there is a rich structure of nonlinear transitions in the dynamics of viscoelastic jets, which transforms the dynamics from simple coiling to the leaping shampoo state via these transitions. 

\section{Discussion and Conclusions}


In this study, we have investigated the effects of viscoelasticity on the dynamics of a continuous jet falling on a rigid plate, specifically relating to the coiling instability and the subsequent dynamics. We have uncovered a rich array of dynamical behavior beyond the initial coiling instability. Our study also presents a unifying picture of seemingly diverse flow situations like coiling, dripping, and the leaping shampoo effect in terms of transitions between different stable regimes as the flow rate and the height of fall are varied.


We have carried out experiments on three different fluids, each with a different viscosity and elasticity. An observation regarding the fluid with lowest viscosity is noteworthy: below a viscosity of $\eta \sim 5$ Pa s, the fluid jet {\emph does not} show buckling and coiling. This is the reason behind our choice for the lowest viscosity to be around $\eta \sim 10$ Pa s, where a steady coiling is obtained after buckling. This observation has not been theoretically understood as yet but it clearly points to a minimum in viscosity for buckling and coiling to occur.  


The regime map for each fluid reveals a number of interesting features. At the lowest flow rate, we always observe dripping beyond a critical height. This critical height is different for each fluid, or the combination ($E_g , Q^*$). This is expected since the jetting-dripping transition or simply the dripping mode in viscous and non-Newtonian fluids itself exhibits a rich, highly nonlinear dynamics \citep{Bas-1,Bas-2}. In the present study we do not undertake a detailed investigation of the jetting-dripping transition. Beyond the lowest flow rate, we encounter a generic behavior of sequence of transitions described above. 

At larger drop heights we encounter the well-known Kaye effect, which still remains a mystery. We conclude that shear-thinning rheology is essential for the onset of Kaye effect, consistent with the observations of Versluis et al. On the other hand, elastic effects are also important for the onset of the Kaye effect. Shear thinning but highly elastic fluids do not exhibit leaping \citep{MV-1,MV-2}, whereas surfactant solutions with $E_g \sim 1$ do show leaping readily or when viscous, gravitational, and elastic forces are in balance. The height at which leaping begins depends on both the flow rate and elasto-gravity number. This observation suggests that leaping requires a delicate balance between a drop in viscosity due to shear, and elasticity.

An important observation regarding the sequence of transitions is that the onset of each state and its duration is strongly dependent on the flow rate. It also depends weakly on the fluid properties, in particular, the elasto-gravity number. In general, as the flow rate becomes higher, the onset of quasi-periodic, irregular, and leaping shampoo behavior occurs at lower heights. These inverse relations may well be power-laws in ($\epsilon,~ Q^*$) space with a weaker dependence on $E_g$. The resolution in height measurement and the limited number of flow rates used in this study prevent us from fitting power-laws to the boundaries of each dynamical state. Although the height between the nozzle and the plate is kept fixed and measured with an accuracy of $1\%$, once the jet impacts the plate, there is always a mound of fluid of varying height throughout the jetting process. The height of the mound is important at low drop heights and hence for the transition from RPC to QPC, and from QPC to MFD states. Although an average height of the mound is subtracted from the nozzle-plate distance, effective error in measured height and actual height can be as large as $10\%$ at low drop heights. 

Two important features are strongly suggested by the present experiments; a nonlinear transition from periodicity via frequency bifurcations and the possible power-law dependence of the regime boundaries in $\epsilon-Q^*$ space. Further insights into the exact nature of nonlinear transitions and the exponents of the power-laws can not be unambiguously determined from the present data. 
 In this situation the role of numerical modeling is critical. Numerical simulations could answer these questions unambiguously. 
So far there have been no analytical modeling studies that deal with continuous jets of viscoelastic fluids beyond the initial coiling instability. Recently a few numerical studies have demonstrated that free surface jets of viscoelastic fluids can be modeled \citep{laso,Mac-1}. Although the goal of these studies was demonstration and effectiveness of the proposed numerical schemes rather than a detailed investigation of the transitions in jet dynamics, such transitions were indeed observed and found to be different compared to Newtonian jets. 

We specially note here a numerical study by Tome et al.  \cite{Mac-2}, who compared  the dynamics of 3D jets of Newtonian and an Oldroyd-B fluid. Amongst other points of comparison, they mention briefly that  while Newtonian jets showed regular coiling, buckled jets of Oldroyd-B fluid began coiling initially, but quickly demonstrated ``apparently chaotic" motion. This is highly significant since we have shown here that there is an order to this ``apparently chaotic'' motion; it comes about via a sequence of transitions and occurs in a small window of the parameter space. It would be highly instructive to apply such modeling schemes to the present problem since the insights gained from such an attempt would be useful for further experimental studies. Such numerical studies, with enough resolution, can settle questions regarding the possible power-law dependence of various regimes boundaries and the exact nonlinear dynamics of the jets all the way from simple coiling to the leaping phenomenon. We hope that the present study stimulates  theoretical and numerical examination of the problem in more detail and help the next phase of experiments. On the experimental front, a next step would be to access higher flow rates and flows in confined geometries which would mimic more closely industrial processes like filling of containers. 

\bibliographystyle{spiebib}
\bibliography{tsm-jet}

\begin{thebibliography}{10}

\bibitem{Ray}
L.~Rayleigh, ``On the capillary phenomena of jets,'' {\em {\rm Proc. R. Soc.
  London }}~{\bf 29}, pp.~71--97, 1879.

\bibitem{GIT}
G.~I. Taylor, ``Instability of jets, threads and sheets of viscous fluid,''
  {\em {\rm Proc. 12th Intl Conf. on Applied Mechanics}} , 1969.

\bibitem{EggersRMP}
J.~Eggers, ``Nonlinear dynamics and breakup of free-surface flows,'' {\em {\rm
  Rev. Mod. Phys. (USA)}}~{\bf 69}(3), pp.~865 -- 929, 1997.

\bibitem{CM-1}
J.~O. Cruicshank and B.~R. Munson, ``Viscous fluid buckling of plane and
  axisymmetric jets,'' {\em {\rm J. Fluid Mech.}}~{\bf 113}, pp.~221--239,
  1981.

\bibitem{oliveira}
M.~S.~N. Oliveira and G.~H. McKinley, ``Iterated stretching and multiple
  beads-on-a-string phenomena in dilute solutions of highly extensible flexible
  polymers,'' {\em {\rm Phys. Fluids}}~{\bf 17}(7), p.~071704, 2005.

\bibitem{GIT-2}
G.~I. Taylor, ``Electrically driven jets,'' {\em {\rm Proc. R. Soc. London,
  Ser. A}}~{\bf 313}(1515), pp.~453--475, 1969.

\bibitem{GT}
R.~W. Griffiths and J.~S. Turner, ``Folding of viscous plumes impinging on
  density or viscosity interface,'' {\em {\rm Geophysical Journal}}~{\bf
  95}(-1), pp.~397--419, 1988.

\bibitem{Barnes-1}
G.~Barnes and R.~Woodcock, ``Liquid rope-coil effect,'' {\em {\rm Am. J. Phys.
  }}~{\bf 26}(4), pp.~205--209, 1958.

\bibitem{Barnes-2}
G.~Barnes and J.~MacKenzie, ``Height of fall versus frequency in liquid
  rope-coil effect,'' {\em {\rm Am. J. Phys. }}~{\bf 27}(2), pp.~112--115,
  1959.

\bibitem{CM-2}
J.~O. Cruicshank and B.~R. Munson, ``The viscous-gravity jet in stagnation
  flow,'' {\em {\rm J. Fluid Eng.}}~{\bf 104}, pp.~360--362, 1982.

\bibitem{CM-3}
J.~O. Cruickshank and B.~R. Munson, ``An energy loss coefficient in fluid
  buckling,'' {\em {\rm Phys. Fluids}}~{\bf 25}(11), pp.~1935--1937, 1982.

\bibitem{CM-4}
J.~O. Cruickshank and B.~R. Munson, ``A theoretical prediction of the fluid
  buckling frequency,'' {\em {\rm Phys. Fluids}}~{\bf 26}(4), pp.~928--930,
  1983.

\bibitem{CM-5}
J.~O. Cruickshank, ``Low-reynolds-number instabilities in stagnating jet
  flows,'' {\em {\rm J. Fluid Mech.}}~{\bf 193}(-1), pp.~111--127, 1988.

\bibitem{Yarin-1}
B.~Tchavdarov, A.~Yarin, and S.~Radev, ``Buckling of thin liquid jets,'' {\em
  {\rm J. Fluid Mech.}}~{\bf 253}(-1), pp.~593--615, 1993.

\bibitem{Yarin-2}
A.~Yarin and B.~M. Tchavdarov, ``Onset of flolding in plane liquid films,''
  {\em {\rm J. Fluid Mech.}}~{\bf 307}(-1), pp.~85--99, 1996.

\bibitem{maha-1}
L.~Mahadevan, W.~S. Ryu, and A.~D.~T. Samuel, ``Fluid rope trick
  investigated,'' {\em {\rm Nature}}~{\bf 392}(-1), p.~140, 1998.

\bibitem{maha-2}
L.~Mahadevan, W.~S. Ryu, and A.~D.~T. Samuel, ``Correction tofluid rope trick
  investigated,'' {\em {\rm Nature}}~{\bf 403}(-1), p.~502, 2000.

\bibitem{Maleki-1}
M.~Maleki, M.~Habibi, R.~Golestanian, N.~M. Ribe, and D.~Bonn, ``Liquid rope
  coiling on a solid surface,'' {\em {\rm Phys. Rev. Lett.}}~{\bf 93},
  p.~214502, Nov 2004.

\bibitem{Habibi-1}
M.~Habibi, M.~Maleki, R.~Golestanian, N.~M. Ribe, and D.~Bonn, ``Dynamics of
  liquid rope coiling,'' {\em {\rm Phys. Rev. E}}~{\bf 74}(6), p.~066306, 2006.

\bibitem{Ribe-1}
N.~M. Ribe, H.~E. Huppert, M.~A. Hallworth, M.~Habibi, and D.~Bonn, ``Multiple
  coexisting states of liquid rope coiling,'' {\em {\rm J. Fluid Mech.}}~{\bf
  555}(-1), pp.~275--297, 2006.

\bibitem{Ribe-2}
N.~M. Ribe, ``Periodic folding of viscous sheets,'' {\em {\rm Phys. Rev. E
  }}~{\bf 68}(3 2), pp.~363051 -- 363056, 2003.

\bibitem{Ribe-4}
N.~M. Ribe, ``A general theory for the dynamics of thin viscous sheets,'' {\em
  {\rm J. Fluid Mech.}}~{\bf 457}(-1), pp.~255--283, 2002.

\bibitem{Habibi-2}
M.~Habibi, N.~M. Ribe, and D.~Bonn, ``Coiling of elastic ropes,'' {\em {\rm
  Phys. Rev. Lett.}}~{\bf 99}(15), p.~154302, 2007.

\bibitem{Ribe-3}
N.~M. Ribe, J.~R. Lister, and S.~Chiu-Webster, ``Stability of a dragged viscous
  thread: Onset of stitching in a fluid-mechanical sewing machine,'' {\em {\rm
  Phys. Fluids}}~{\bf 18}(12), p.~124105, 2006.

\bibitem{Morris}
S.~W. Morris, J.~H.~P. Dawes, N.~M. Ribe, and J.~R. Lister, ``Meandering
  instability of a viscous thread,'' {\em {\rm Phys. Rev. E}}~{\bf 77}(6),
  p.~066218, 2008.

\bibitem{NH}
S.~ichiro Nagahiro and Y.~Hayakawa, ``Bending-filament model for the buckling
  and coiling instability of a viscous rope,'' {\em {\rm Phys. Rev. E.}}~{\bf
  78}(-1), p.~025302(R), 2008.

\bibitem{Matovich}
M.~A. Matovich and J.~R. Pearson, ``Spinning a molten thread,'' {\em {\rm Ind.
  and Eng. Chem. Fundamentals}}~{\bf 8}(3), pp.~512--520, 1969.

\bibitem{MM}
F.~W. Kroesser and S.~Middleman, ``Viscoelastic jet stability,'' {\em {\rm
  AIChE Journal}}~{\bf 15}(3), pp.~383--387, 1969.

\bibitem{BFW-1}
S.~Bechtel, M.~Forest, and D.~Bogy, ``A one-dimensional theory for viscoelastic
  fluid jets, with application to extrudate swell and draw-down under
  gravity,'' {\em {\rm J. Non-Newtonian Fluid Mech.}}~{\bf 21}(3), pp.~273 --
  308, 1986.

\bibitem{BFW-2}
S.~Bechtel, J.~Cao, and M.~Forest, ``Practical application of a higher order
  perturbation theory for slender viscoelastic jets and fibers,'' {\em {\rm J.
  Non-Newtonian Fluid Mech.}}~{\bf 41}(3), pp.~201 -- 73, 1992.

\bibitem{BFW-3}
M.~Forest and Q.~Wang, ``Change-of-type behavior in viscoelastic slender jet
  models,'' {\em {\rm Theoretical and Computational Fluid Dynamics}}~{\bf
  2}(1), pp.~1 -- 25, 1990.

\bibitem{AK}
A.~Kaye, ``A bouncing liquid stream,'' {\em {\rm Nature}}~{\bf 197}(-1),
  pp.~1001--1002, 1963.

\bibitem{CF}
A.~A. Collyer and P.~J. Fisher, ``The kaye effect revisited,'' {\em {\rm
  Nature}}~{\bf 261}(-1), pp.~682--683, 1976.

\bibitem{Vers}
M.~Versluis, C.~Blom, D.~van~der Meer, K.~van~der Weele, and D.~Lohse,
  ``Leaping shampoo and the stable kaye effect,'' {\em {\rm J. Stat. Mech.:
  Theory and Experiment}} (07), 2006.

\bibitem{GHM-1}
G.~H. McKinley, ``Dimensionless groups for understanding free surface flows of
  complex fluids,'' {\em {\rm Society of Rheology Bulletin}} (-1), pp.~6--9,
  2005.

\bibitem{zak}
M.~Zak, ``Dynamics of liquid films and thin jets,'' {\em {\rm SIAM J. App.
  Math.}}~{\bf 37}(2), pp.~276--289, 1979.

\bibitem{MV-1}
M.~Varagnat, {\em (Th - Masters Thesis) Instabilities of jets of non-Newtonian
  fluids impacting a plate}.
\newblock PhD thesis, Massachusetts Institute of Technology, May 2008.

\bibitem{MV-2}
M.~Varagnat, T.~S. Majmudar, W.~Hartt, and G.~H. McKinley, ``The folding motion
  of axi-symmetric jets of worm-like micellar solutions,'' 2009.

\bibitem{Bas-1}
B.~Ambravaneswaran, H.~Subramani, S.~Phillips, and O.~Basaran,
  ``Dripping-jetting transitions in a dripping faucet,'' {\em {\rm Phys. Rev.
  Lett.}}~{\bf 93}(3), pp.~034501--1, 2004.

\bibitem{Bas-2}
H.~Subramani, H.~K. Yeoh, R.~Suryo, Q.~Xu, B.~Ambravaneswaran, and O.~Basaran,
  ``Simplicity and complexity in a dripping faucet,'' {\em {\rm Phys.
  Fluids}}~{\bf 18}(3), pp.~32106 -- 1, 2006.

\bibitem{laso}
M.~Picasso, A.~Bonito, and M.~Laso, ``Numerical simulation of 3d viscoelastic
  flows with free surfaces,'' {\em {\rm J. Comp. Phys.}}~{\bf 215}(2), pp.~691
  -- 716, 2006.

\bibitem{Mac-1}
M.~Tome, N.~Mangiavacchi, J.~Cuminato, A.~Castelo, and S.~McKee, ``A finite
  difference technique for simulating unsteady viscoelastic free surface
  flows,'' {\em {\rm J. Non-Newtonian Fluid Mech.}}~{\bf 106}(2-3), pp.~61 --
  106, 2002.

\bibitem{Mac-2}
M.~Tome, A.~Castelo, V.~Ferreira, and S.~McKee, ``A finite difference technique
  for solving the oldroyd-b model for 3d-unsteady free surface flows,'' {\em
  {\rm J. Non-Newtonian Fluid Mech.}}~{\bf 154}(2-3), pp.~179 -- 206, 2008.

\end{thebibliography}
 



\clearpage
\section{Supplementary Figures}
\begin{figure}[bthp!]
\centering
\vspace{5pt}\
\includegraphics[width=450pt]{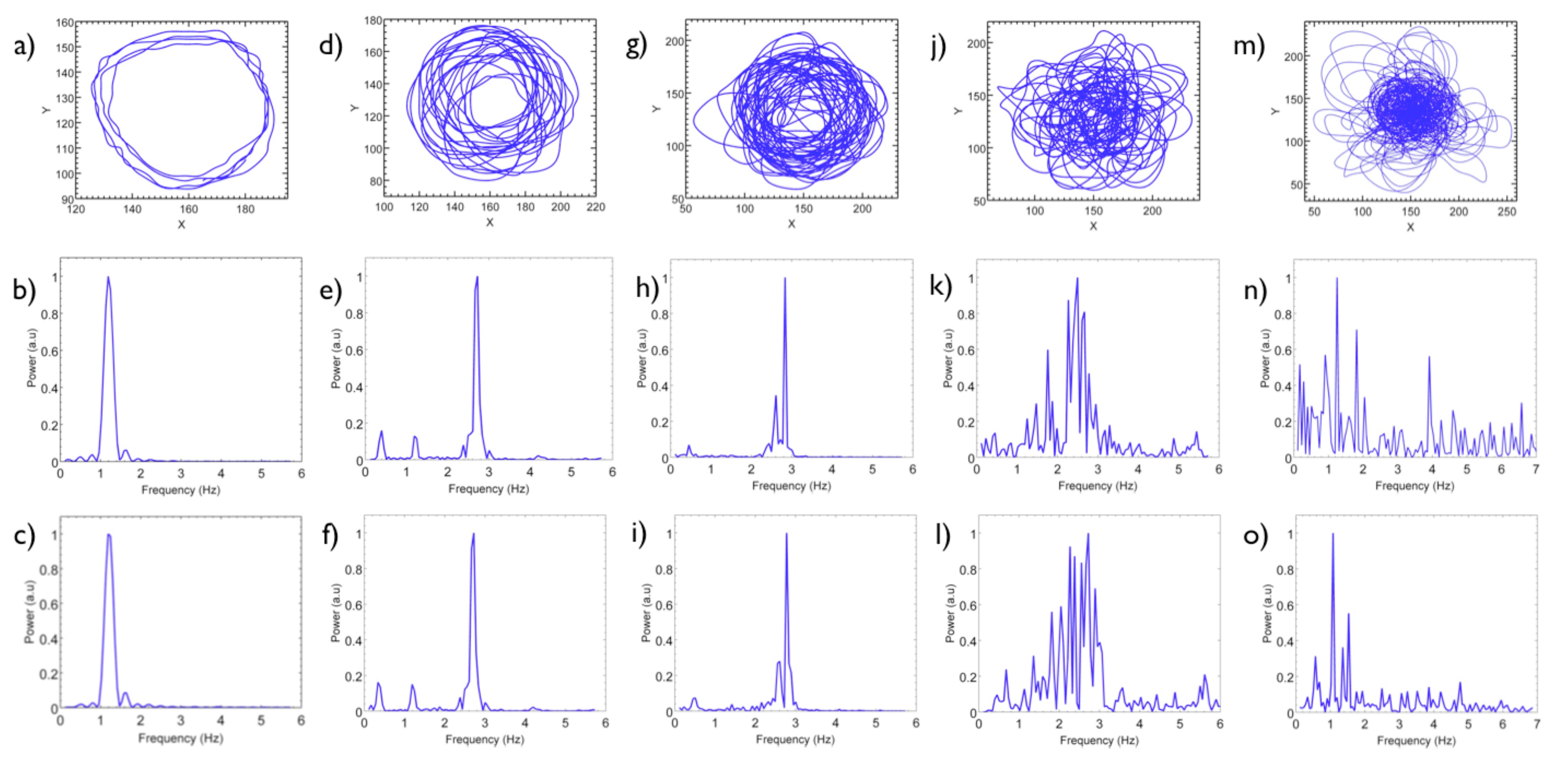}
\caption{Supplementary Figure 1. Trajectory tracking and power spectra (linear scale): Top row shows the trajectories of the jets of fluid S2 at a flow rate of $Q^* = 0.8 \times 10^{-4}$, for different heights. The power spectra of the $X(t)$ component are in the middle row, and the power spectra of the $Y(t)$ components are in the bottom row. a) Trajectory of the RPC state, at $H/d = 8.33$. b,c)  A single frequency $f_0=1.2 {\rm Hz}$ is clearly visible. d) The trajectory of the QPC regime at $H/d = 12.5$. The orbits are slightly elliptical and the jet exhibits coiling and precession. e,f) Power spectra of the $X(t),Y(t)$ components, respectively. The peak frequency at the previous height $f_0$ is still present but at a much lower amplitude. There are two additional frequencies, $f_1 = 2.72 {\rm Hz}$ and $f_2 = 0.40 {\rm Hz}$. g) The trajectory of another type of QPC dynamics showing coiling and folding states, which always appears after the rotation-precession dynamics. h,i) The power spectra of the components. j) The trajectory corresponding to the multi-frequency state. The jet exhibits complex coiling, folding and meandering dynamics giving rise to multiple frequency components. k,l) The power spectra corresponding to the $X(t)$ and $Y(t)$ components, respectively. The multi-frequency nature of the spectra is evident. m) The trajectory  at the onset of the leaping-shampoo state. Localized coiling and folding states are interrupted by large amplitude leaps in the $X-Y$ plane. n,o) The power spectra now have many frequencies and the power is distributed more uniformly among many frequencies.} 
\label{fig:Supplementary Figure 1}
\end{figure}
\clearpage

\begin{figure}[bthp!]
\centering
\vspace{5pt}\
\includegraphics[width=450pt]{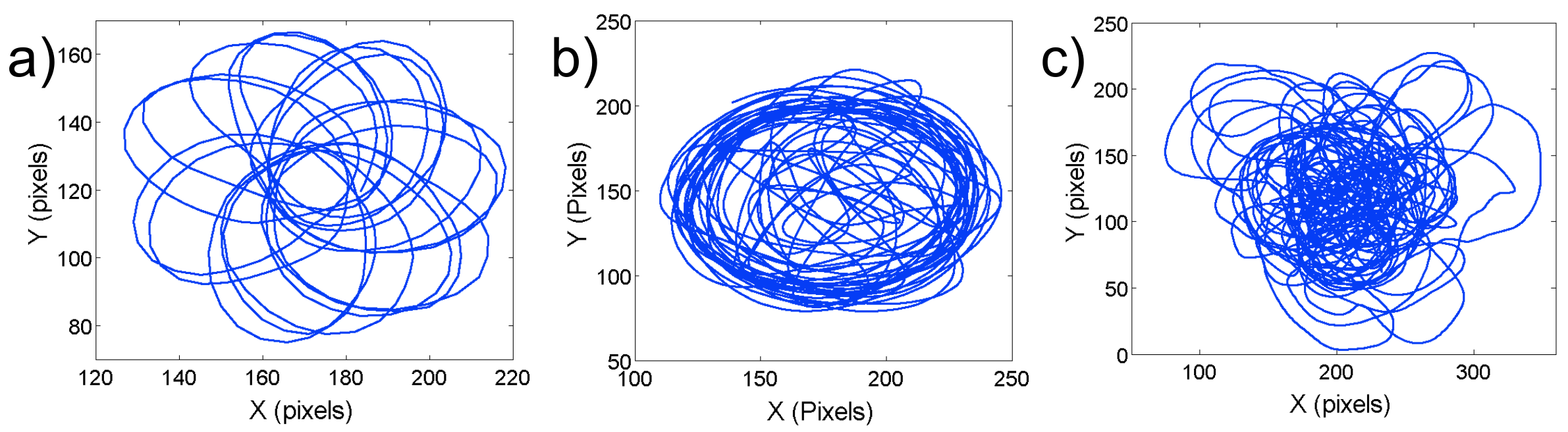}
\caption{Supplementary Figure 2. Representative trajectories with two or more frequencies. a) Fluid S2: $Q^* = 0.4 \times 10^{-4}$, $\epsilon = 12.5$; the rotation-precession mode is clearly visible giving rise to petal-like trajectories in two dimensions. b) Fluid S1: $Q^* = 2.3 \times 10^{-4}$, $\epsilon = 16.7$; coiling states are intermittently disrupted by folding dynamics or linear strokes across the circular trajectory, sometimes via a ``figure-of-eight''. c) Fluid S3: $Q^* =  3.8 \times 10^{-5}$, $\epsilon = 33.3$; coiling, folding, and long meandering loops in the trajectory, each with its distinct frequency resulting in multi-frequency dynamics.}
\label{fig:Supplementary Figure 2}
\end{figure}



\end{document}